\documentclass[aps,prb,amsmath,amssymb,reprint,superscriptaddress,preprintnumbers,showpacs,intlimits,longbibliography]{revtex4-1}
\pdfoutput=1
\bibpunct{[}{]}{,}{n}{}{}

\usepackage{bm,latexsym,mathrsfs,enumerate}
\usepackage{mathtools}
\usepackage[breaklinks=true,unicode=true,urlcolor = blue,colorlinks = true,citecolor = blue,linkcolor = blue]{hyperref}
\usepackage{graphicx}
\renewcommand{\vec}[1]{\bm{#1}}
\begin{document}

\title{Geometry-induced motion of magnetic domain walls in curved nanostripes}

\author{Kostiantyn V. Yershov}
\email[Corresponding author: ]{yershov@bitp.kiev.ua}
\affiliation{Bogolyubov Institute for Theoretical Physics of the National Academy of Sciences of Ukraine, 03143 Kyiv, Ukraine}
\affiliation{National University of Kyiv-Mohyla Academy, 04655 Kyiv, Ukraine}

\author{Volodymyr P. Kravchuk}
\email{vkravchuk@bitp.kiev.ua}
\affiliation{Bogolyubov Institute for Theoretical Physics of the National Academy of Sciences of Ukraine, 03143 Kyiv, Ukraine}
\affiliation{Leibniz-Institut f\"ur Festk\"orper- und Werkstoffforschung, IFW Dresden, Dresden D-01171, Germany}

\author{Denis D. Sheka}
\email{sheka@knu.ua}
\affiliation{Taras Shevchenko National University of Kyiv, 01601 Kyiv, Ukraine}

\author{Oleksandr V. Pylypovskyi}
\email{engraver@knu.ua}
\affiliation{Taras Shevchenko National University of Kyiv, 01601 Kyiv, Ukraine}

\author{Denys Makarov}
\email{d.makarov@hzdr.de}
\affiliation{Helmholtz-Zentrum Dresden-Rossendorf~e.~V., Institute of Ion Beam Physics and Materials Research, 01328 Dresden, Germany}

\author{Yuri Gaididei}
\email{ybg@bitp.kiev.ua}
\affiliation{Bogolyubov Institute for Theoretical Physics of the National Academy of Sciences of Ukraine, 03143 Kyiv, Ukraine}

\date{May 11, 2018}

%
%%%%%%%%%%%%%%%%%%%%%%%%%%%%%%%%%%%%%%%%%%%%%%%%%%%%%%%%%%%%%%%%%%%%%
%
%         ABSTRACT
%
%%%%%%%%%%%%%%%%%%%%%%%%%%%%%%%%%%%%%%%%%%%%%%%%%%%%%%%%%%%%%%%%%%%%%
%
\begin{abstract}
	Dynamics of topological magnetic textures are typically induced externally by, e.g.~magnetic fields or spin/charge currents. Here, we demonstrate the effect of the internal-to-the-system geometry-induced motion of a domain wall in a curved nanostripe. Being driven by the gradient of the curvature of a biaxial stripe, transversal domain walls acquire remarkably high velocities of up to $100$ m/s and do not exhibit any Walker-type speed limit. We pinpoint that the inhomogeneous distribution of the curvature-induced Dzyaloshinskii--Moriya interaction is a driving force for the motion of a domain wall. Although we showcase our approach on the specific Euler spiral geometry, the approach is general and can be applied to a wide class of geometries.
\end{abstract}

% 75.30.Et	Exchange and superexchange interactions (see also 71.70.Gm Exchange interactions)
% 75.40.Mg	Numerical simulation studies
% 75.60.Ch	Domain walls and domain structure 
% 75.78.Cd	Micromagnetic simulations
% 75.78.Fg	Dynamics of domain structures
\maketitle

%%%%%%%%%%%%%%%%%%%%%%%%%%%%%%%%%%%%%%%%%%%%%%%%%%%%%%%%%%%%%%%%%%%%%
%
%         INTRODUCTION
%
%%%%%%%%%%%%%%%%%%%%%%%%%%%%%%%%%%%%%%%%%%%%%%%%%%%%%%%%%%%%%%%%%%%%%
\section{Introduction}\label{s:introduction}

The deterministic manipulation of magnetic textures, e.g. domain walls~(DWs) and skyrmions, in magnetic stripes is a key practical task to realize high-speed, high-density, low-power, and non-volatile  memory devices~\cite{Parkin08,Xu08a,Fert13,Parkin15}. Typically, the motion of DWs is realized externally by applying a magnetic field~\cite{Thiaville02a} or electric current~\cite{Slonczewski96,Bazaliy98,Zhang04}. The main hurdle on the way towards achieving high translational speed for DWs is the appearance of the Walker limit~\cite{Schryer74,Thiaville06,Mougin07}, which imposes a maximum value of the driving force (magnetic field, current density) for translational motion. Several approaches  have been proposed to achieve a high-speed translational DW motion: usage of antiferromagnetically coupled magnetic nanowires~\cite{Yang15b,Meng16}, application of spin currents perpendicular to the wire \cite{Khvalkovskiy09}, application of spin-orbit torques for chiral DWs \cite{Emori13,Ryu13}. In spite of the fact that maximum value of the driving force is zero for the case of head-to-head (tail-to-tail) DWs in uniaxial wires, one can realize DW motion with a high and constant velocity in the precessional regime with a uniform rotation of the DW phase~\cite{Yan10}. Furthermore, it was shown that the curvature has drastic influence on the Walker limit \cite{Landeros10, Otalora12a,Otalora12,Otalora13,Hertel16,Yan11a}. For example, under certain conditions, the Walker limit can be suppressed in nanotubes~\cite{Hertel16,Yan11a}. Alternative way to achieve the motion of DWs is based on the deformation of the DW structure and known as automotion~\cite{Chauleau10,Nikonov14,Richter16,Mawass17}. This type of motion can be realized relying on the coordinate-dependent cross-sectional area of a nanostripe~\cite{Richter16,Mawass17}, nucleation of DWs with inertial motion~\cite{Nikonov14} or transformation of DWs from the transversal to the vortex~\cite{Chauleau10} by short current pulses.

Here, we propose a concept of geometry-induced motion of topological defects in a curved nanostripe. We demonstrate that a DW performs a translational motion under the influence of the gradient of the stripe curvature. As we do not observe a transition to the precessional regime of motion in the case of a biaxial as well as uniaxial nanowire, the geometry-induced motion is free of a Walker-type speed limit. We pinpoint that the inhomogeneous distribution of the curvature-induced Dzyaloshinskii--Moriya interaction~(DMI) driven by the exchange~\cite{Gaididei14,Sheka15} acts as a driving force for the motion of transversal DWs in curved nanostrips. We propose a general approach valid for a wide class of geometries. The analytical results are confirmed by means of micromagnetic simulations.

%%%%%%%%%%%%%%%%%%%%%%%%%%%%%%%%%%%%%%%%%%%%%%%%%%%%%%%%%%%%%%%%%%%%%
%
%         MAIN PART
%
%%%%%%%%%%%%%%%%%%%%%%%%%%%%%%%%%%%%%%%%%%%%%%%%%%%%%%%%%%%%%%%%%%%%%
\section{The Model and Equations of motion}\label{s:general_results}

%======================================================================================================================
%														FIGURE 1
%======================================================================================================================
\begin{figure}[b]
	\includegraphics[width=\columnwidth]{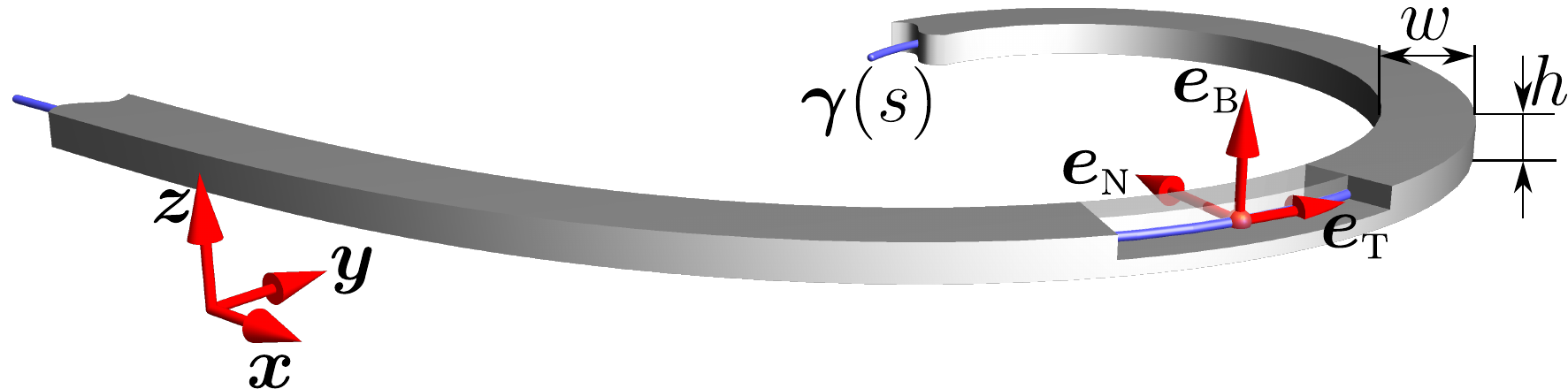}
	\caption{\label{fig:geometry}%
		(Color online) Geometry and notations: we consider a one-dimensional curved biaxial ferromagnet of thickness $h$ and width $w$ with the easy-axis $\vec{e}_\textsc{t}$ and easy-plane $\textsc{tn}$ anisotropies.}
\end{figure} 
%======================================================================================================================
%======================================================================================================================

We consider a flat narrow curved ferromagnetic stripe of a rectangular cross-section whose thickness and width are small enough to ensure the magnetization uniformity along a wire cross-section. The stripe length is substantially larger than the transversal dimensions. Thus, the magnetization is described by the continuous and normalized function $\vec{m}=\vec{M}/M_s=\vec{m}(s,t)$, where $M_s$ is the saturation magnetization, $s$ is the arc length coordinate, and $t$ denotes time. Such a stripe can be parametrized in the following way $\vec{r}(s,\xi_1,\xi_2)=\vec{\gamma}(s)+\xi_1\vec{e}_\textsc{n}(s)+\xi_2\vec{e}_\textsc{b}(s)$. Here, the three-dimensional radius vector $\vec{r}$ defines the space domain, occupied by the stripe, $\vec{\gamma}(s)=\gamma_x(s)\hat{\vec{x}}+\gamma_y(s)\hat{\vec{y}}$ is a two-dimensional vector, which lies within the $xy$-plane and determines the stripe center line, see Fig.~\ref{fig:geometry}. The parameters $\xi_1\in\left[-w/2,w/2\right]$ and $\xi_2\in\left[-h/2,h/2\right]$ are coordinates in the transversal cross-section $w\geq h$, see Fig.~\ref{fig:geometry}.

The magnetic properties of a narrow ferromagnetic stripe are described using a model of a classical ferromagnetic wire with a biaxial anisotropy. The easy-axis is tangential to the stripe central line $\vec{\gamma}$ whereas the easy-plane coincides with the stripe plane~(\textsc{tn}-plane). The magnetic energy of the stripe normalized by $4\pi M_s^2$ reads
\begin{equation}\label{eq:energy_model}
\mathcal{E}=\mathcal{S}\int_{-\infty}^{\infty}\left[\mathscr{E}_\text{ex}-k_a\left(\vec{m}\cdot\vec{e}_\textsc{t}\right)^2+k_p\left(\vec{m}\cdot\vec{e}_\textsc{b}\right)^2\right]\mathrm{d}s.
\end{equation}
Here, $\mathcal{S}=hw$ is the cross-section area. The first term in~\eqref{eq:energy_model} is the exchange energy density  $\mathscr{E}_\text{ex}=\ell^2\sum_{i=x,y,z}\left(\partial_i \vec{m}\right)^2$ with $\ell=\sqrt{A/\left(4\pi M_s^2\right)}$ being the exchange length where $A$ is the exchange constant. The last two terms in~\eqref{eq:energy_model} determine the anisotropy energy density. Vectors $\vec{e}_\textsc{t}$ and $\vec{e}_\textsc{b}$ are unit vectors along the anisotropy axes, which are assumed to be oriented along the tangential and binormal directions~\cite{Note1}. Constants $k_a=K_a/\left(4\pi M_s^2\right)+k_a^\text{ms}$ and $k_p=K_p/\left(4\pi M_s^2\right)+k_p^\text{ms}$ are dimensionless anisotropy constants of easy-tangential and easy-plane types, respectively, with $K_a>0$ and $K_p>0$ being magneto-crystalline anisotropy constants. Terms  $k_a^\text{ms}$ and $k_p^\text{ms}$ arise from the magnetostatic contribution. It is known~\cite{Hillebrands06,Porter04,Aharoni98} that the magnetostatic energy of a straight and uniformly magnetized stripe with rectangular cross-section can be reduced to the effective shape anisotropy~\cite{Aharoni98} with constants
\begin{equation}\label{eq:anisotropy}
\begin{split}
k_a^\text{ms}&=\frac{\frac{1-\delta^2}{2\delta}\ln\left(1+\delta^2\right)+\delta\ln\delta+2\arctan\frac{1}{\delta}}{2\pi},\\
k_p^\text{ms}&=\frac{1}{2}-2 k_a^\text{ms},\quad \delta=w/h\geq1.
\end{split}
\end{equation}
For thin, narrow, and curved stripes~(ribbons) the approximation of the shape anisotropy is used also for inhomogeneous magnetization states~\cite{Gaididei17a}, including domain walls~\cite{Mougin07}. In the limit case of square~($w/h=1$) or circular cross-sections, the magnetostatic-shape-induced anisotropy coefficients~\eqref{eq:anisotropy} are simplified to $k_a^\text{ms}=1/4$ and $k_p^\text{ms}=0$, which is a well known result~\cite{Hillebrands06,Kravchuk14c} including the case of curvilinear wires~\cite{Slastikov12}.

The energy density~\eqref{eq:energy_model} in terms of the angular parametrization $\vec{m}=\vec{e}_\textsc{t}\cos\theta+\vec{e}_\textsc{n}\sin\theta\cos\phi+\vec{e}_\textsc{b}\sin\theta\sin\phi$ has the following form
\begin{equation}\label{eq:energy_angular}
\begin{split}
\mathscr{E}=\ell^2\left \{\left(\theta'+\kappa\cos\phi\right)^2+\left[\phi'\sin\theta-\kappa\cos\theta\sin\phi\right]^2\right\}\\
+\sin^2\theta\left(k_a+k_p\sin^2\phi\right),
\end{split}
\end{equation}
where the first term corresponds to the exchange energy density in the curved wire~\cite{Sheka15} with $\kappa$ being a curvature of the $\vec{\gamma}$. In \eqref{eq:energy_angular} it is taken into account that a flat wire has zero torsion.

%======================================================================================================================
%														FIGURE 2
%======================================================================================================================
\begin{figure}
	\includegraphics[width=0.95\columnwidth]{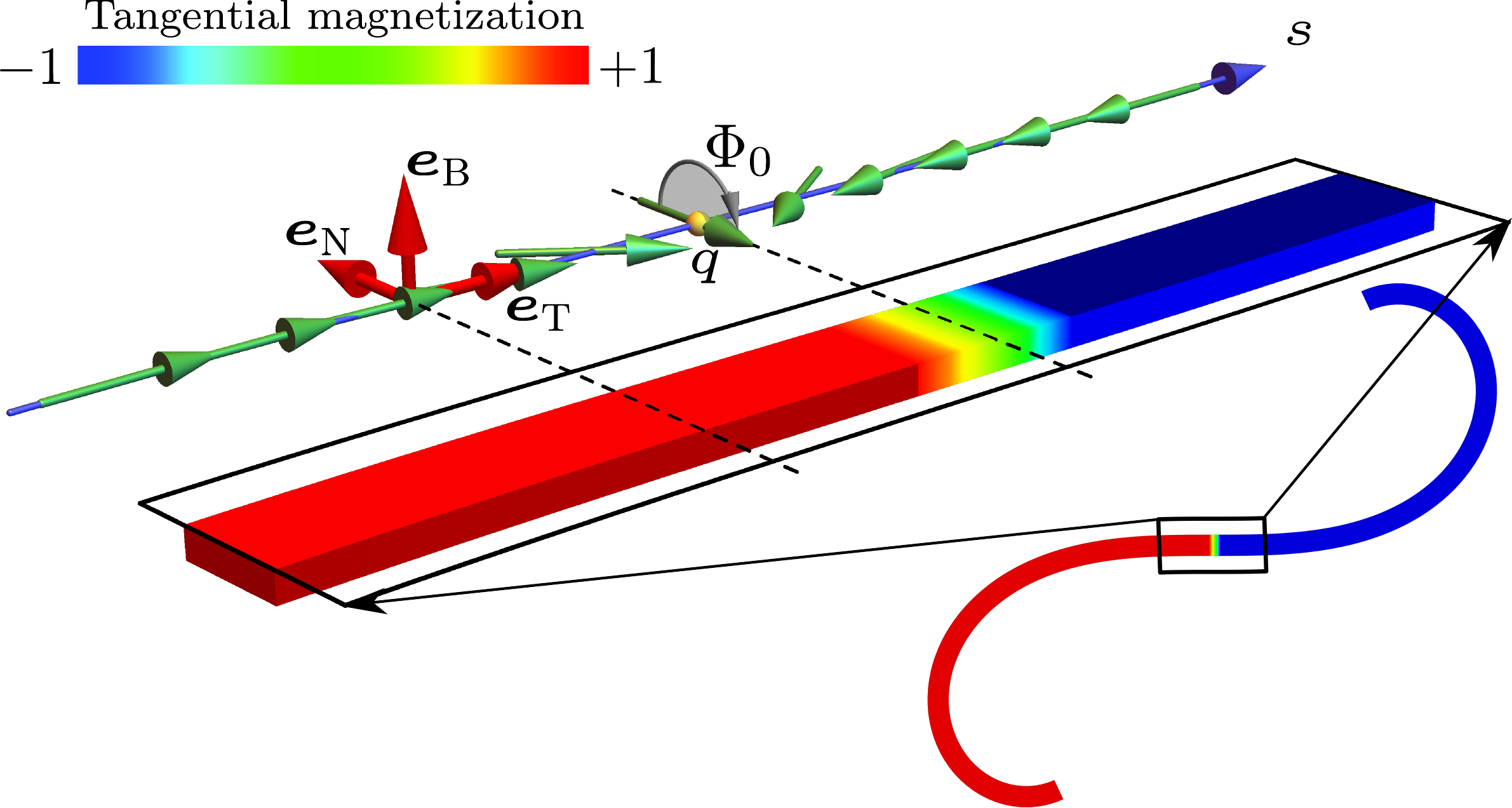}
	\caption{\label{fig:q_phi_scheme}%
		(Color online) Illustration of a one-dimensional head-to-head DW ($p=+1$ and $\mathcal{C}=-1$) geometry in a stripe with shape of an Euler spiral described by two collective coordinates: the DW position $q$ and phase~$\Phi$. Red axes determine curvilinear basis; green arrows and color scheme determine the magnetization distribution in the stripe obtained by means of \texttt{Nmag} micromagnetic simulations. Simulation is performed for the Permalloy stripe with $h=5$ nm, $w=15$ nm, and $\chi\ell^2=2\times 10^{-4}$ in an overdamped regime~($\alpha=0.5$).}
\end{figure} 
%======================================================================================================================
%======================================================================================================================

%======================================================================================================================
%														FIGURE 3
%======================================================================================================================
\begin{figure*}
	\includegraphics[width=\textwidth]{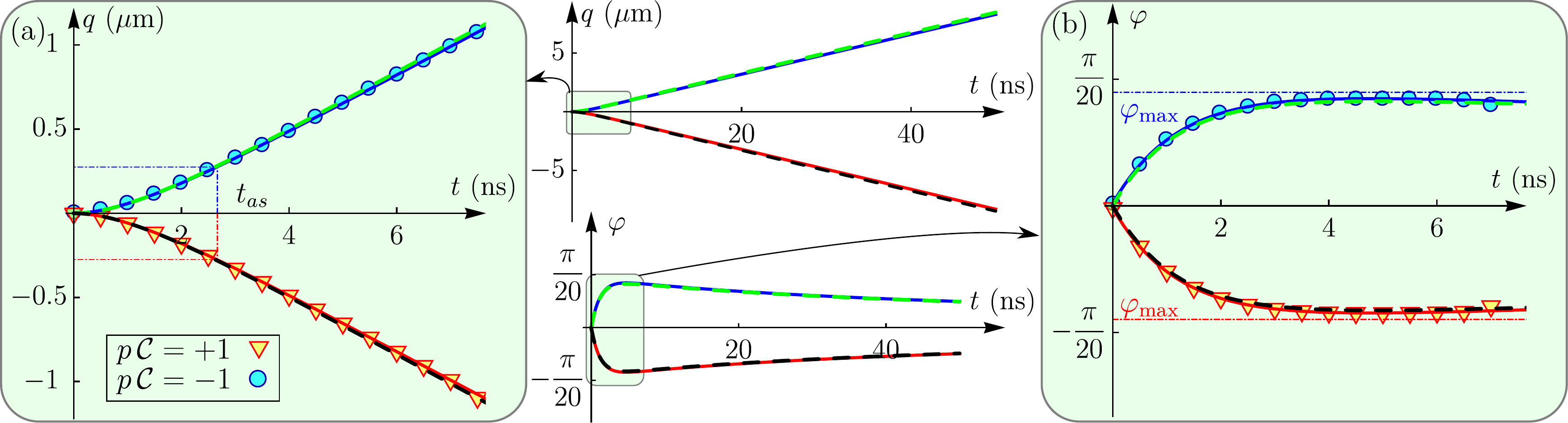}
	\caption{\label{fig:q_phi}%
		(Color online) Typical behavior of the DW position (a) and phase deviation (b) for a head-to-head DW ($p=+1$). Geometrical parameters of the stripe are as follows: $\chi\ell^2=2\times10^{-4}$, $w=15$ nm, $h=5$ nm.  Solid and dashed lines correspond to solutions of the collective variables equations \eqref{eq:dynamic_equation} and predictions \eqref{eq:phi_expression}-\eqref{eq:q_expression} of the linearized model, respectively. Asymptotic time calculated accordingly to~\eqref{eq:asympt_time} for $\varepsilon=0.9$. Symbols show the results of \texttt{Nmag} micromagnetic simulations. Central inset demonstrates the comparison of solutions of the collective variables equations \eqref{eq:dynamic_equation} and predictions \eqref{eq:phi_expression}-\eqref{eq:q_expression} of long time dynamics. In all cases $\alpha=0.01$ and $K_a=K_p=0$.}
\end{figure*} 
%======================================================================================================================
%======================================================================================================================

To analyze the dynamics of a DW in a curved magnetic stripe we use a collective variable approach based on the $q-\Phi$ model~\cite{Slonczewski72,Malozemoff79}
\begin{equation}\label{eq:q_phi_model}
\cos\theta=-p\tanh\frac{s-q}{\Delta},\quad \phi=\Phi.
\end{equation}
Here, $\{q,\Phi\}$ are time-dependent conjugated collective variables, which determine the DW position and phase, respectively, see Fig.~\ref{fig:q_phi_scheme}; $\Delta$ is a DW width; $p$ is a topological charge, which determines the DW type: head-to-head~($p=+1$) or tail-to-tail~($p=-1$). The model~\eqref{eq:q_phi_model} coincides with the exact DW solution for a rectilinear wire~($\kappa'\equiv0$). In the following, the curvature is considered as a small perturbation, which results in the DW drift while keeping the form~\eqref{eq:q_phi_model} unchanged. The analysis is carried out in the approximation linear with respect to the curvature and its gradient: $\kappa\ell/\sqrt{k_a}\ll1$ and $\kappa'\ell^2/k_a\ll1$.
 
Substituting the Ansatz~\eqref{eq:q_phi_model} into~\eqref{eq:energy_angular} and performing integration over the arc length $s$, we obtain the energy of a DW in a curved stripe in the form~(up to an additive constant and quadratic terms with respect to $\kappa$)
\begin{equation}\label{eq:energy_qPhi}
\frac{\mathcal{E}}{2\mathcal{S}}\approx\frac{\ell^2}{\Delta}+\Delta k_a +\Delta k_p \sin^2{\Phi}+p\pi\kappa(q)\ell^2\cos\Phi,
\end{equation}
where the condition $\kappa\Delta\ll1$ was applied when integrating~\eqref{eq:energy_angular}. First three terms in \eqref{eq:energy_qPhi} determine the competition of the isotropic exchange and anisotropy contributions, while the forth term originates from the curvilinear-geometry-induced~DMI driven by the exchange~\cite{Gaididei14,Sheka15}.

In terms of the collective variables, the equations of motion take a form (see Appendix~\ref{app:q_phi})
\begin{equation}\label{eq:dynamic_equation}
\begin{split}
\frac{\alpha}{\Delta}\dot{q}+p\dot{\Phi}&=-p\pi\omega_0\ell^2\frac{\partial\kappa(q)}{\partial q}\cos\Phi,\\
p\dot{q}-\alpha\Delta\dot{\Phi}&=-p\pi\omega_0\ell^2\kappa(q)\sin\Phi+k_p\Delta\omega_0\sin2\Phi,
\end{split}
\end{equation}
where $\omega_0$ is a characteristic time scale of the system, $\alpha$ is a damping parameter. The DW width is assumed to be a slaved variable~\cite{Hillebrands06,Landeros10}, i.e., $\Delta(t)\equiv\Delta\left[\Phi(t)\right]=\ell/\sqrt{k_a+k_p\sin^2\Phi}$. From~\eqref{eq:dynamic_equation} it follows that the gradient of the curvature is a driving force for DWs. The physical origin of this force is the curvilinear-geometry-induced~DMI driven by the exchange.

The ground state in a curved wire~($\kappa'\neq0$) cannot be strictly tangential: the magnetization vector $\vec{m}$ deviates from the tangential direction by an angle $\vartheta\approx\chi\ell^2/k_a$ (for the case $\kappa'\equiv\chi$) in the \textsc{tn}-plane. As shown in Appendix~\ref{app:rotated}, this effect results in the modification of Eqs.~\eqref{eq:dynamic_equation} up to corrections in curvature and its gradient of the second order of magnitude. Therefore, the carried out analysis of the DW motion in the approximation linear with respect to the curvature and its gradient is valid.

\section{Domain Wall Dynamics in Euler Spirals}\label{s:linear_curvature}

In the following we apply the general $q-\Phi$ equations of motion~\eqref{eq:dynamic_equation} for a particular case of an Euler spiral~\cite{Lawrence14}, also known as Cornu spiral or clothoid, see Fig.~\ref{fig:q_phi_scheme}. The equation for the central  line of such a stripe has the form
\begin{equation}\label{eq:euler_spiral}
\vec{\gamma}(s)=\hat{\vec{x}}\int_{0}^{s}\cos\left(\frac{\chi}{2}\zeta^2\right)\mathrm{d}\zeta+\hat{\vec{y}}\int_{0}^{s}\sin\left(\frac{\chi}{2}\zeta^2\right)\mathrm{d}\zeta.
\end{equation}
The curvature in this case is a linear function of the arc length coordinate $\kappa\left(s\right)=\chi s$ with $\chi$ being the gradient of the curvature. It is necessary to mention that we are interested in the stripes of a finite width $w$. Therefore, to avoid an overlap between the neighboring windings of the spiral, the distance between them must be bigger than the stripe width $w$. The minimal distance between windings is determined by the condition~$\kappa w\ll 1$.

Using a small-angle approximation for the DW phase  $\varphi=\Phi-\Phi_0$~($\varphi\ll 1$, see Appendix~\ref{app:calc_details}), we obtain the asymptotic expression for the wall velocity
\begin{equation}\label{eq:dw_velocity}
V= -p\,\mathcal{C}\pi\Delta_0\omega_0\frac{\chi\ell^2}{\alpha}, 
\end{equation}
where $\mathcal{C}=\cos\Phi_0=\pm1$ with $\Phi_0$ being the initial DW phase, $\Delta_0=\ell/\sqrt{k_a}$. In the following it will be shown that the initial value of the DW phase coincides with $\Phi(t\to\infty)$. Therefore, we can interpret $\mathcal{C}$ as the DW magnitochirality~\cite{Kim14}. Remarkably, the DW velocity~\eqref{eq:dw_velocity} is similar to the well known expression~\cite{Thiaville05} $V^u=u\beta/\alpha$ in magnetic biaxial stripes caused by the Zhang--Li mechanism~\cite{Bazaliy98,Zhang04}, where $\beta$ is a nonadiabatic spin-transfer parameter. Current-induced translational DW motion takes place only if $u<u_\textsc{w}$, where $u_\textsc{w}$ being Walker current~\cite{Thiaville05,Mougin07}. However, for the case of a geometry-induced motion, a Walker-limit-like effect of the transition to the precessional regime does not appear and the DW demonstrates a high-speed translational motion without any external driving. The DW behavior~\eqref{eq:dw_velocity} is also similar to the dynamics of bubbles in a gradient magnetic field~\cite{Malozemoff79}. Still, in our case the DW moves in the direction of the gradient of the curvature, while bubbles are displaced in the perpendicular direction to the gradient of the field. 

%======================================================================================================================
%														FIGURE 4
%======================================================================================================================
\begin{figure}
	\includegraphics[width=0.9\columnwidth]{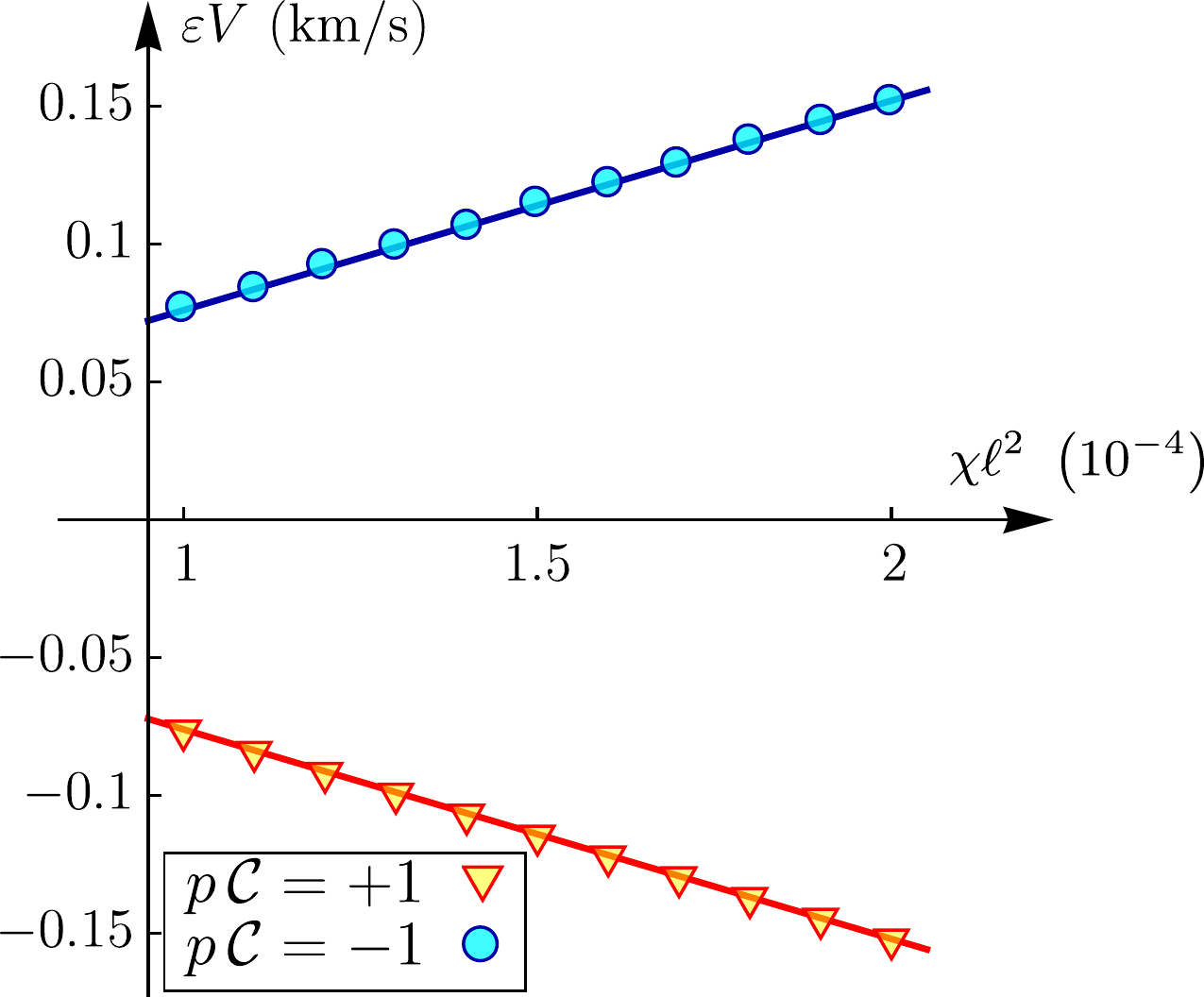}
	\caption{\label{fig:velocity_stripe}%
		(Color online) DW velocity as a function of the gradient of the curvature. Solid lines correspond to the prediction \eqref{eq:dw_velocity} with the asymptotic parameter $\varepsilon=0.9$. Symbols show the the DW velocity at moment $t_{as}$ obtained from \texttt{Nmag} micromagnetic simulations for the stripe with $w=15$~nm, $h=5$~nm, and $K_a=K_p=0$~(for details see Appendix~\ref{app:simulations}).}
\end{figure} 
%======================================================================================================================
%======================================================================================================================

The DW velocity~\eqref{eq:dw_velocity} is independent of the easy-plane anisotropy coefficient. However, this coefficient determines the time needed for the DW velocity to reach the asymptotic value~\eqref{eq:dw_velocity}. This time can be estimated as~(for details see Appendix~\ref{app:calc_details})
\begin{equation}\label{eq:asympt_time}
t_{as}\approx -\frac{1+\alpha^2}{\omega_0}\frac{\ln\left[\left(1+\alpha^2\right)\left(1-\varepsilon\right)\right]}{2\alpha k_p+\chi\ell^2},
\end{equation}
where $\varepsilon<1$ is an asymptotic parameter. Additionally, the easy-plane anisotropy determines the maximal DW phase value $\Phi_\text{max}=\Phi_0+\varphi_\text{max}$ with $\varphi_\text{max}\approx-\mathcal{C}\pi\chi\ell^2/\left(2\alpha k_p\right)\ll1$ and $k_p>0$. At long timescale, the change of the DW phase can be written as~(for details see Appendix~\ref{app:calc_details})
\begin{equation}\label{eq:long_phase}
\Phi\approx\Phi_0+p\frac{\Delta_0}{\alpha V t},
\end{equation}
which results in the condition $\Phi\left(t\to\infty\right)=\Phi_0$. A typical time evolution of the DW position $q(t)$ and phase deviation $\varphi(t)$ is shown in Fig.~\ref{fig:q_phi}.

In the no-damping approximation (see Appendix~\ref{app:zero_damp}), the DW motion differs from~\eqref{eq:dw_velocity} and \eqref{eq:long_phase}. For the case of zero damping, the DW phase deviation reaches the value $\varphi(\alpha=0)\to-\mathcal{C}\pi/2$, while the velocity of the DW increases exponentially with time within the considered model. Note that for the velocities larger than the minimal phase velocity of magnons, the model should be revised by including the Cherenkov-like effect~\cite{Yan11a}.

We checked the theoretically predicted velocities for the DW motion~\eqref{eq:dw_velocity} by micromagnetic simulations of magnetically soft stripes with material parameters of Permalloy~\cite{Note3} using \texttt{Nmag} code~\cite{Fischbacher07}, see Figs.~\ref{fig:q_phi}-\ref{fig:velocity_stripe} and Appendix~\ref{app:simulations} for details. The numerics agrees well with the analytical prediction~\eqref{eq:dw_velocity}.

The resulting DW velocity as a function of the gradient of the curvature is plotted in Fig.~\ref{fig:velocity_stripe}. The DW velocity increases almost linearly with the gradient of the curvature. The direction of the DW motion depends on the sign of the product of the topological charge $p$, magnitochirality $\mathcal{C}$ of the DW, and sign of the gradient of the curvature $\chi$ ($\dot{q}\propto p\,\mathcal{C}\chi$), see Figs.~\ref{fig:q_phi}(a) and~\ref{fig:velocity_stripe}. In Fig.~\ref{fig:q_phi}(a)  the DW position is shown as a function of time for walls with different sign of the product of the topological charge and magnitochirality: for $p\,\mathcal{C}=-1$, the DW moves in the direction of the increasing curvature. For the case $p\,\mathcal{C}=+1$, DW moves in the opposite direction~(in both cases $\chi>0$)~\cite{Note4}.

\section{Conclusion}\label{sec:consl}

In conclusion, we predict the effect of geometry-induced motion of a DW in a curved nanostripe: DWs are driven by the gradient of the stripe curvature without any external stimuli, see Eqs.~\eqref{eq:dynamic_equation} and~\eqref{eq:dw_velocity}. The physical origin of the driving force is the curvature-induced~DMI driven by the exchange~\cite{Gaididei14,Sheka15}. Geometry-induced motion results in a high-speed translational motion of the DW position without transition into the precessional regime. The latter effect can be interpreted as a curvature-induced suppression of the Walker limit. We show that the direction of the DW motion is determined by the product of the DW magnitochirality, topological charge, and gradient of the curvature $\dot{q}\propto p\,\mathcal{C}\chi$, see Eq.\eqref{eq:dw_velocity}, while the change of the DW phase at long timescales results in $\Phi-\Phi_0\propto 1/q\left(t\right)$. Additionally, it is necessary to mention that the coefficient of the easy-plane anisotropy determines the time~\eqref{eq:asympt_time}, which is needed for the DW to reach the asymptotic velocity~\eqref{eq:dw_velocity}.

Linear model of the curvature-induced DW motion~\eqref{eq:dynamic_equation} can be used also for a 3D wires with small torsion. The latter contributes to the negligibly small quadratic corrections. However, role of the torsion becomes significant, when the spin-torques are applied~\cite{Yershov16}.

\section*{Acknowledgments}\label{s:acknowledgments}
K.~Y., O.~P., and D.~Sh. acknowledge Helmholtz-Zentrum Dresden-Rossendorf, where part of this work was performed, for kind hospitality. K. Y. acknowledges a financial support from DAAD (Code No. 91618879). D.~Sh. and O.~P. acknowledge the support from the Alexander von Humboldt Foundation (Research Group Linkage Programme). D.~M. acknowledges the support via the BMBF project GUC-LSE (FKZ: 01DK17007) and German Science Foundation (DFG) Grant MA 5144/9-1. The present work was partially supported by the Program of Fundamental Research of the Department of Physics and Astronomy of the National Academy of Sciences of Ukraine (Project No. 0116U003192).
\appendix

\section{Details of the $q-\Phi$ model}\label{app:q_phi}

Magnetization dynamics of this system is studied by means of phenomenological Landau--Lifshitz--Gilbert equations
\begin{equation}\label{eq:LLG_ang}
-\sin\theta\dot{\theta}=\omega_0\frac{\delta\mathcal{E}}{\delta\phi}+\alpha\sin^2\theta\dot{\phi},\quad \sin\theta\dot{\phi}=\omega_0\frac{\delta\mathcal{E}}{\delta\theta}+\alpha\dot{\theta},
\end{equation}
where constant $\alpha$ is a Gilbert damping coefficient, overdot  indicates the time derivative, frequency $\omega_0=4\pi\gamma_0M_s$ determines the characteristic timescale of the system with $\gamma_0$ being the gyromagnetic ratio.

The equations of motion \eqref{eq:LLG_ang} are the Lagrange--Rayleigh equations
\begin{equation}\label{eq:Lagrange_Rayleigh}
\frac{\delta\mathcal{L}}{\delta X_i}-\frac{\mathrm d}{\mathrm d t}\frac{\delta\mathcal{L}}{\delta\dot{X}_i}=\frac{\delta\mathcal{F}}{\delta\dot{X}_i},\qquad X_i\in\{\theta,\,\phi\}
\end{equation}
for Lagrange function~\cite{Doering48}
\begin{equation}\label{eq:Lagrange}
\mathcal{L}=-\mathcal{S}\int\limits_{-\infty}^{\infty}\phi\sin\theta\dot{\theta}\mathrm{d}s-\omega_0\mathcal{E}
\end{equation}
and dissipative function~\cite{Thiaville02,Gilbert04}
\begin{equation}\label{eq:diss_func}
\mathcal{F}=\frac{\alpha}{2}\mathcal{S}\int\limits_{-\infty}^{\infty}\left[\dot{\theta}^2+\sin^2\theta\dot{\phi}^2\right]\mathrm{d}s.
\end{equation}
Substituting the Ansatz \eqref{eq:q_phi_model} into \eqref{eq:Lagrange} and \eqref{eq:diss_func} and performing the integration over the arc length $s$ we obtain the Lagrange and dissipative functions in the form
\begin{equation}\label{eq:Lag_diss_qPhi}
\mathcal{L}=2p\mathcal{S}\Phi\dot{q}-\omega_0\mathcal{E},\quad \mathcal{F}=\frac{\alpha}{\Delta}\mathcal{S}\left[\dot{q}^2+\left(\Delta\dot{\Phi}\right)^2+c\dot{\Delta}^2\right],
\end{equation}
where $c=\pi^2/12$. Substituting \eqref{eq:Lag_diss_qPhi} into the Lagrange--Rayleigh equations \eqref{eq:Lagrange_Rayleigh} results in the equations of motion
\begin{equation}\label{eq:qPhiDelta_motion}
	\begin{split}
		\frac{\alpha}{\Delta}\dot{q}+p\dot{\Phi}&=-p\pi\omega_0\ell^2\frac{\partial\kappa(q)}{\partial q}\cos\Phi,\\
		p\dot{q}-\alpha\Delta\dot{\Phi}&=-p\pi\omega_0\ell^2\kappa(q)\sin\Phi+k_p\Delta\omega_0\sin2\Phi,\\
		c\frac{\alpha}{\omega_0}\dot{\Delta}&=\frac{\ell^2}{\Delta}-\Delta\left(k_a+k_p\sin^2\Phi\right).
	\end{split}
\end{equation}
The third equation in~\eqref{eq:qPhiDelta_motion} shows that $\Delta$ relaxes towards its equilibrium value $\Delta_0 = \ell/\sqrt{k_a+ k_p \sin^2\Phi}$. The characteristic time $t_\text{relax}$ of this relaxation is proportional to the damping $t_\text{relax}\propto \alpha/\omega_0$~\cite{Hillebrands06}. Usually $\alpha\ll1$, therefore one can conclude that the DW width is a slave variable $\Delta(t)=\Delta\left[\Phi(t)\right]$ and DW dynamics can be described by the set~\eqref{eq:dynamic_equation} with the equilibrium DW width $\Delta=\Delta_0$.

\section{Equations of motion in a rotated reference frame}\label{app:rotated}

The static magnetization distribution of the system is determined by minimum of the energy~\eqref{eq:energy_angular}. Minimization of~\eqref{eq:energy_angular} with respect to $\phi$ results in the solution~$\phi=\phi_0=\{0,\pi\}$, while minimization with respect to $\theta$ results in an inhomogeneous Sine-Gordon equation~(for details see Ref.~\onlinecite{Yershov15b})
\begin{equation}\label{eq:theta_Eq}
\frac{\ell^2}{k_a}\theta''-\sin\theta\cos\theta=-\kappa'\frac{\ell^2}{k_a}\cos\phi_0.
\end{equation}
This equation has a homogeneous solution for $\kappa'=\chi$~(the case of an Euler spiral)
\begin{equation}\label{eq:theta_hom}
\theta_0=\frac{1}{2}\arcsin\left(2\chi\frac{\ell^2}{k_a}\cos\phi_0\right).
\end{equation}
This means that the magnetization is not oriented along the $\textsc{tnb}$ basis.

Now we will rotate the reference frame in a local rectifying surface by the angle $\vartheta=\frac{1}{2}\arcsin\left(2\chi\frac{\ell^2}{k_a}\cos\varPhi\right)$ using a unitary transformation
\begin{equation}\label{eq:transformation}
\begin{split}
\vec{m}=U\tilde{\vec{m}},\quad \tilde{\vec{m}}=U^{-1}\vec{m},\quad\tilde{\vec{m}}=\{m_1,m_2,m_3\}^\textsc{t},\\
U=\left(\begin{matrix}
\cos\vartheta&-\text{sgn}(\cos\varPhi)\sin\vartheta&0\\
\text{sgn}(\cos\varPhi)\sin\vartheta&\cos\vartheta&0\\
0&0&1
\end{matrix}\right).
\end{split}
\end{equation}
After this transformation, the energy density can be written as
\begin{equation}\label{eq:energy_1}
\begin{split}
\mathscr{E}=\ell^2\left[|\tilde{\vec{m}}'|^2+2\chi s\left(m_1m_2'-m_1'm_2\right)+\left(\chi s\right)^2\left(1-m_3^2\right)\right]\\
-k_a\left[\cos\vartheta m_1-\text{sgn}(\cos\phi)\sin\vartheta m_2\right]^2+k_pm_3^2
\end{split}
\end{equation}
Using the angular parametrization $\tilde{\vec{m}}=\{\cos\Theta,\sin\Theta\cos\varPhi,\sin\Theta\sin\varPhi\}$, the energy density~\eqref{eq:energy_1} can be written in the form
\begin{equation}\label{eq:energy_2}
\begin{split}
\mathscr{E}=\ell^2\biggl[\Theta'^2+\varPhi'^2\sin^2\Theta+\left(\chi s\right)^2\left(1-\sin^2\varPhi\sin^2\Theta\right)\\
+\chi s\left(2\Theta'\cos\varPhi-\varPhi'\sin 2\Theta\sin\varPhi\right)\biggr]+k_p\sin^2\varPhi\sin^2\Theta\\-k_a\left[\cos\vartheta \cos\Theta-\text{sgn}(\cos\varPhi)\sin\vartheta \cos\varPhi\sin\Theta\right]^2.
\end{split}
\end{equation}

Static form of  Landau--Lifshitz--Gilbert equations~\eqref{eq:LLG_ang} read $\delta\mathcal{E}/\delta\Theta=0$ and $\delta\mathcal{E}/\delta\varPhi=0$. Taking into account the energy density~\eqref{eq:energy_2} one obtains the following set of equations:
\begin{widetext}
\begin{subequations}\label{eq:stat_eq}
	\begin{equation}\label{eq:stat_eq_theta}
	\begin{split}
	\sin\Theta \cos\Theta \left[k_a \left(\cos^2\vartheta-\cos^2\varPhi \sin^2\vartheta\right)+k_p \sin^2\varPhi\right]+\frac{1}{2} k_a \cos 2\Theta \cos\varPhi \sin 2\vartheta\text{sgn}(\cos\varPhi)\\
	+\ell^2 \left[-\chi^2 s^2\sin\Theta \cos\Theta \sin^2\varPhi-\chi\cos\varPhi-\Theta''+\varPhi'\sin\Theta\left(\varPhi'\cos\Theta+2 \chi s \sin\Theta \sin\varPhi\right)\right]=0
	\end{split}
	\end{equation}
	\begin{equation}\label{eq:stat_eq_phi}
	\begin{split}
	-k_a \cot\Theta \sin\varPhi \sin 2\vartheta\text{sgn}(\cos\varPhi)+&k_a\sin 2\varPhi \sin^2\vartheta+\sin 2\varPhi \left(k_p-\ell^2 \chi^2 s^2\right)\\
	-2 \ell^2 \left(2 \Theta'\varPhi' \cot\Theta+ \varPhi''\right)&+2 \ell^2 \chi \sin\varPhi \left(\cot\Theta-2 s \Theta'\right)=0
	\end{split}
	\end{equation}
\end{subequations}
\end{widetext}

Equation~\eqref{eq:stat_eq_phi} has a homogeneous solution $\varPhi=\varPhi_0=\{0,\pi\}$. Substitution of this solution into~\eqref{eq:stat_eq_theta} results in
\begin{equation}\label{eq:theta_rot}
\sin 2 (\Theta +\vartheta )-2 \frac{\ell^2}{k_a} \Theta''=\sin 2\vartheta,
\end{equation}
where we use $\vartheta=\frac{1}{2}\arcsin\left(2\chi\frac{\ell^2}{k_a}\cos\varPhi_0\right)$. Equation~\eqref{eq:theta_rot} has a homogeneous solution $\Theta=\Theta_0=\{0,\pi\}$. However, this equation has not static solution of a DW type with boundary conditions $\cos\Theta\left(\pm\infty\right)=\mp p$.

In the following we will modify $q-\Phi$ model~\eqref{eq:q_phi_model} for angles $\Theta$ and $\varPhi$
\begin{equation}\label{eq:q_phi_model_rot}
\cos\Theta=-p\tanh\frac{s-q}{\Delta},\quad \varPhi=\Phi.
\end{equation}
By inserting Ansatz~\eqref{eq:q_phi_model_rot} into the energy density~\eqref{eq:energy_2} and integrating over the arc length $s$, we obtain
\begin{equation}\label{eq:energy_rotated}
\begin{split}
\frac{\mathcal{E}^\textsc{dw}}{2\mathcal{S}}\approx \frac{\ell^2}{\Delta}+\Delta\left(k_p+\frac{1}{2}k_a-\chi^2q^2\right)\sin^2\Phi\\+\frac{1}{4}\Delta k_a\cos 2\vartheta\left(3+\cos 2\Phi\right)+p\pi\ell\chi q\cos\Phi.
\end{split}
\end{equation}

Equations of motion~\eqref{eq:dynamic_equation} will be modified in the following way
\begin{equation}\label{eq:dynamic_equation1}
\begin{split}
\frac{\alpha}{\Delta}\dot{q}+p\dot{\Phi}&=-p\pi\omega_0\ell^2\chi\cos\Phi+2\omega_0\ell^2 \Delta \chi^2 q \sin^2\Phi,\\
p\dot{q}-\alpha\Delta\dot{\Phi}&=-p\pi\omega_0\ell^2\chi q\sin\Phi+\Delta\omega_0\sin2\Phi\\
\times&\biggl[k_p - \ell^2 \chi^2 q^2 + \frac{1}{4}\left(k_a-\sqrt{k_a^2-4\chi^2\ell^4}\right)\biggr],\\
c\frac{\alpha}{\omega_0}\dot{\Delta}=&\frac{\ell^2}{\Delta}-\Delta\Biggl[\left(k_p+\frac{k_a}{2}-\chi^2 q^2\right)\sin^2\Phi\\
&+\frac{1}{2}\sqrt{k_a^2-4\chi^2q^2}\left(2-\sin^2\Phi\right)\Biggr].
\end{split}
\end{equation}
Equations of motion \eqref{eq:dynamic_equation1} coincide with \eqref{eq:dynamic_equation} and \eqref{eq:qPhiDelta_motion} up to corrections in $\chi$ of the second order of magnitude.

\section{Details of the DW motion in an Euler spiral}\label{app:calc_details}

In the limit case of constant curvature $\kappa'=0$, static Eqs.~\eqref{eq:LLG_ang} have a solution of a DW with the phase $\Phi_0=\pi$ and $\Phi_0=0$. Therefore, to analyze the dynamics of DW in stripes with the non-zero gradient of the curvature $\kappa'\neq 0$ and $\kappa'\ell^2/k_a\ll1$ we will use small-angle approximation $\varphi(t)=\Phi(t)-\Phi_0$, which is valid for stripes with in-plane magnetization. Using this approximation, equations of motion \eqref{eq:dynamic_equation} can be written in the form
\begin{equation}\label{eq:qPhi_phi_deviation}
\begin{split}
&\frac{\alpha}{\Delta_0}\dot{q}+p \dot{\varphi}=-p\mathcal{C} \pi\omega_0\ell^2\chi,\\
&p \dot{q}-\alpha  \Delta_0  \dot{\varphi}=-p\mathcal{C}\pi\omega_0\ell^2\chi q\varphi+2k_p\Delta_0\omega_0\varphi,
\end{split}
\end{equation}
where $\mathcal{C}=\cos\Phi_0$  with $\Phi_0$ being the initial DW phase.

The first equation in \eqref{eq:qPhi_phi_deviation} can be simply integrated, while the second can be written as
\begin{equation}\label{eq:phi_deviation}
\begin{split}
\left(1+\alpha^2\right)\dot{\varphi}=-\mathcal{C}\pi\omega_0\ell^2\chi-\pi^2\omega^2_0\ell^4\chi^2\varphi t\\
-\left(2k_p+p\mathcal{C}\pi\ell^2\chi\frac{q_0}{\Delta_0}\right)\alpha\omega_0\varphi,
\end{split}
\end{equation}
where we keep only linear terms with respect to $\varphi$. Here, $q_0$ is the initial DW position. Equation~\eqref{eq:phi_deviation} has the following solution
	\begin{equation}\label{eq:phi_expression}
	\begin{split}
	\varphi(t)=-\mathcal{C}\sqrt{\frac{2}{1+\alpha^2}}\Biggl\{F\left(\mathfrak{D}+\frac{\pi\ell^2\chi}{\sqrt{2\left(1+\alpha^2\right)}}\omega_0t\right)	-F\left(\mathfrak{D}\right)\\
	\times\exp\left[-\left(2\mathfrak{D}+\frac{\pi\ell^2\chi}{\sqrt{2\left(1+\alpha^2\right)}}\omega_0t\right)\frac{\pi\ell^2\chi}{\sqrt{2\left(1+\alpha^2\right)}}\omega_0t\right]\Biggr\},
	\end{split}
	\end{equation}
where $\mathfrak{D}=\alpha\left(2k_p+p\,\mathcal{C}\pi\ell^2\chi\frac{q_0}{\Delta_0}\right)/\left[\pi\ell^2\chi\sqrt{2\left(1+\alpha^2\right)}\right]$, $F(x)=\exp\left(-x^2\right)\int_{0}^{x}\exp\left(y^2\right)\mathrm{d}y$ is a Dawson's integral~\cite{NIST10}. DW phase deviation~\eqref{eq:phi_expression} allows to obtain the following equation for the DW position
\begin{equation}\label{eq:q_expression}                                                                                                                                               
q(t)=q_0-Vt-p\frac{\Delta_0}{\alpha}\varphi(t),
\end{equation}
where $V=-p\,\mathcal{C}\pi\Delta_0\omega_0\chi\ell^2/\alpha$ is an asymptotic DW velocity. The time needed for the DW velocity to reach the asymptotic value $V$ can be found as a solution of the equation $\dot{q}(t_\text{as})=\varepsilon V$, where $\varepsilon<1$ is an asymptotic parameter. 

The precise analysis of the DW phase deviation~\eqref{eq:phi_expression} results in the following expression $\varphi\approx p\Delta_0/\left(\alpha V t\right)$ at long timescale.

\section{No-damping approximation for the DW motion in an Euler spiral}\label{app:zero_damp}

Here, we consider a limiting case of zero damping ($\alpha=0$). In this case, equations of motion can be written as 
\begin{equation}\label{eq:de_no_damping}
\begin{split}
\dot{\Phi}&=-\pi\omega_0\ell^2\chi\cos\Phi,\\
\dot{q}&=-\pi\omega_0\ell^2\chi q\sin\Phi+pk_p\Delta_0\omega_0\sin2\Phi.
\end{split}
\end{equation}
The solution of Eqs.~\eqref{eq:de_no_damping} are 
\begin{equation}\label{eq:no_damping_sol}
\begin{split}
\Phi=&\Phi_0-\mathcal{C}\frac{\pi}{2}+2\arctan e^{-\pi\omega_0\ell^2\chi t},\\
q=&\left(q_0-p\,\mathcal{C}\frac{k_p\Delta}{\pi\ell^2\chi}\tanh p\pi\omega_0\ell^2\chi t\right)\cosh p\pi\omega_0\ell^2\chi t.
\end{split}
\end{equation}

The behavior of first equation in \eqref{eq:no_damping_sol} at long timescale corresponds to $\varphi\to-\mathcal{C}\pi/2$, while the second equation shows that the DW position changes with an exponential law with time limitation $\omega_0t<\left(\pi\ell^2\chi\right)^{-1}\ln\left[2\pi\ell^2/\left(k_p\Delta^2\right)\right]$.

\section{Numerical simulations}\label{app:simulations}

%======================================================================================================================
%														FIGURE 5
%======================================================================================================================
\begin{figure}
	\includegraphics[width=0.85\columnwidth]{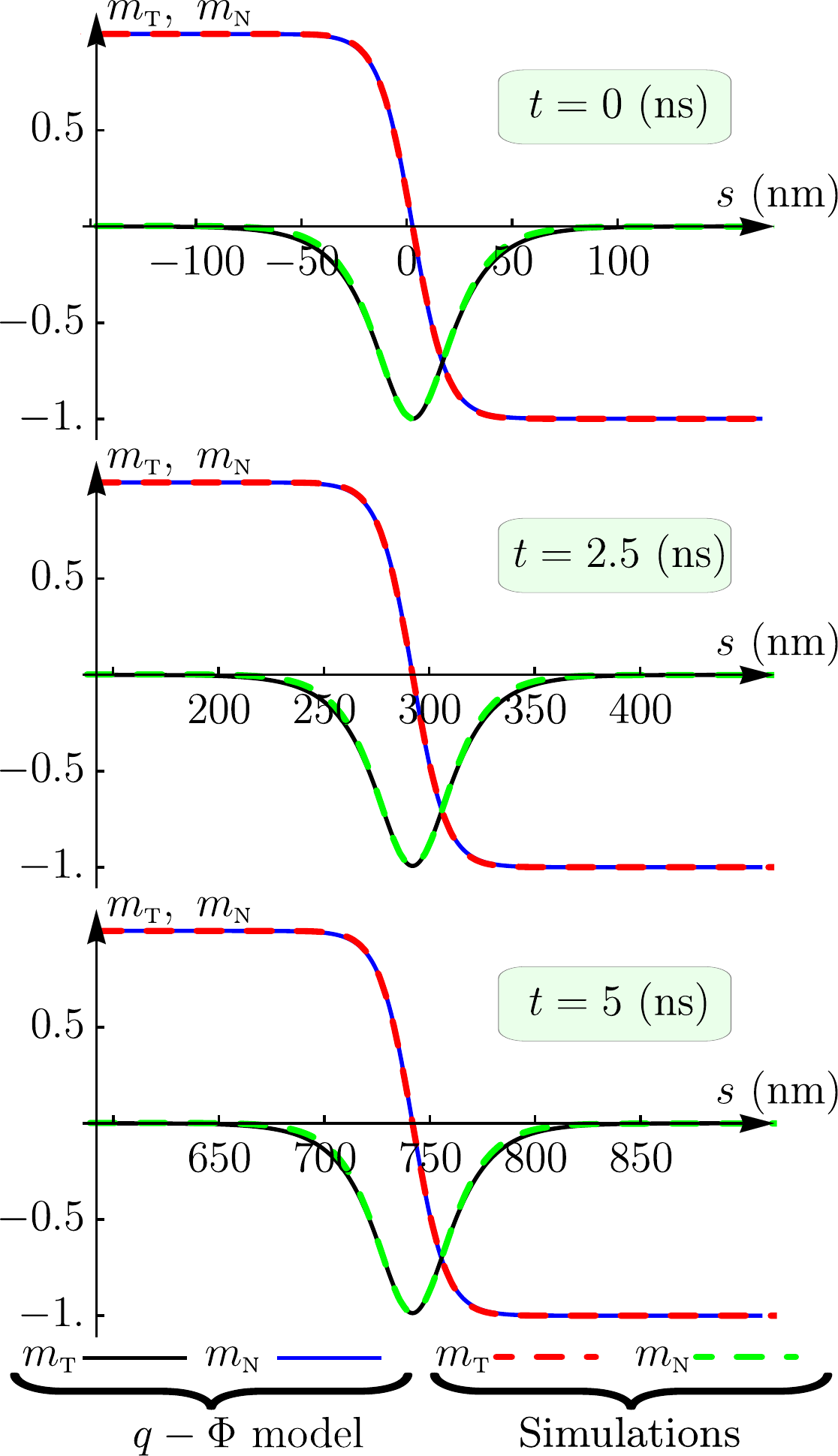}
	\caption{\label{fig:dw_profile}%
		(Color online) Comparison of the magnetization components $m_\textsc{t}=\vec{m}\cdot\vec{e}_\textsc{t}$ and $m_\textsc{n}=\vec{m}\cdot\vec{e}_\textsc{n}$ obtained by means of \texttt{Nmag} micromagnetic simulations and from the Ansatz \eqref{eq:q_phi_model}: $\Phi_0=\pi$. Simulations are performed for a Permalloy stripe with the gradient of the curvature $\chi\ell^2=2\times 10^{-4}$ and damping $\alpha=0.01$. In the Ansatz~\eqref{eq:q_phi_model} the DW width $\Delta_0=\ell/\sqrt{k_a}$ was used.}
\end{figure} 
%======================================================================================================================
%======================================================================================================================

To verify our analytical results, we perform numerical micromagnetic simulations of the Landau--Lifshitz--Gilbert equation utilizing the \texttt{Nmag} code~\cite{Fischbacher07}. We restrict ourselves to the case of magnetically soft material. Therefore only two magnetic interactions are taken into account, namely the exchange and magnetostatic contributions.

We consider stripes whose central line is determined by \eqref{eq:euler_spiral}. In simulations we use material parameters of Permalloy~\cite{Note3}. The dimensions of a stripe are fixed for all studied cases~(thickness $h=5$ nm, width $w=15$ nm, and length $L=2\ \mu$m), while the curvature is varied in the range $\chi\ell^2\in\left[1,2\right]\times10^{-4}$. An irregular tetrahedral mesh with a cell size about of 2.75 nm is used.

The numerical experiment consists of two steps. First, we relaxed the DW with certain values of topological charge $p$ and DW magnitochirality $\mathcal{C}$ in a curved stripe in an overdamped regime~($\alpha=0.5$), see Fig.~\ref{fig:q_phi_scheme} and Fig.~\ref{fig:dw_profile} with $t=0$ ns.  After relaxation, the magnetization dynamics are simulated for a typically used value of the damping coefficient~$\alpha=0.01$. 

To determine the values of $q$ and $\Phi$,  we extract the curvilinear magnetization components $m_\textsc{t}=\vec{m}\cdot\vec{e}_\textsc{t}$, $m_\textsc{n}=\vec{m}\cdot\vec{e}_\textsc{n}$, and $m_\textsc{b}=\vec{m}\cdot\vec{e}_\textsc{b}$ from the simulation data and apply fitting with the Ansatz \eqref{eq:q_phi_model}. Namely, the position $q$ is determined as a fitting parameter for the function $m_\textsc{t}(s)=-p\tanh[(s-q)/\Delta]$, then the phase is determined from the equation $\tan\Phi=m_\textsc{b}(q)/m_\textsc{n}(q)$. A typical behavior of the DW position and phase is plotted in Fig.~\ref{fig:q_phi}. The tangential component of the DW profile is determined with the relative error smaller than $10^{-5}$\%.

The velocity of DW is calculated as $\dot{q}\left(t_{as}\right)$, where $\dot{q}$ is extracted from mictomagnetic simulations, $t_{as}$ is calculated with Eq.~\eqref{eq:asympt_time} for an asymptotic parameter $\varepsilon=0.9$.

\end{document}